\documentstyle[12pt]{article}
\def\tend{\mathop{\to}}

\def\lim{\mathop{\rm {lim}}}
minus0.1667em \baselineskip2\bigskipamount
\begin{document}
\large
\begin{center}
\bigskip
{\bf Renormalization of equations governing nucleon
dynamics \\
and nonlocality of the NN interaction. }\\
\bigskip
{\rm Renat Kh.Gainutdinov and Aigul A.Mutygullina}\\\bigskip \it
Department of Physics,
Kazan State University,\\
18 Kremlevskaya St, Kazan 420008,
Russia\\
E-mail: Renat.Gainutdinov@ksu.ru
\end{center}
\section*{\bf Abstract}
\vskip 1.5em

\parindent=2.5em

{\rm\par We discuss the problem of renormalization of dynamical
equations which arises in an effective field theory description of
nuclear forces. By using a toy model of the separable NN potential
leading to logarithmic singularities in the Born series, we show
that renormalization gives rise to nucleon dynamics which is
governed by a generalized dynamical equation with a
nonlocal-in-time interaction operator.

PACS: 21.45.+v, 02.70.-c, 24.10.-i

Keywords: Effective field theory; Nucleon dynamics;
Nonlocal-in-time interaction.

 \newpage
\parindent=2.5em

 Understanding how
nuclear forces emerge from the fundamental theory of quantum
chromodynamics (QCD) is one of the most important problem of
quantum physics. To study hadron dynamics at scales where QCD is
strongly coupled, it is useful to employ effective field theories
(EFT's) [1] being an invaluable tool for computing physical
quantities in the theories with disparate energy scales. In order
to describe low energy processes involving nucleons and pions, all
possible interaction operators consistent with the symmetries of
QCD are included in an effective Lagrangian of an EFT. However
such a Lagrangian leads to ultraviolet (UV) divergences that must
be regulated and a renormalization scheme defined. A fundamental
difficulty in an EFT description of nuclear forces is that they
are nonperturbative, so that an infinite series of Feynman
diagrams must be summed. Summing the relevant diagrams is
equivalent to solving a Schr{\"o}dinger equation. However, an EFT
yields graphs which are divergent, and gives rise to a singular
Schr{\"o}dinger potential. For this reason N-nucleon potentials
are regulated and renormalized couplings are defined [2].
Nevertheless, a renormalization procedure does not lead to the
potentials satisfying the requirements of ordinary quantum
mechanics, and consequently after renormalization nucleon dynamics
is not governed by the Schr{\"o}dinger equation. This raises the
question of what kind of equation governs nucleon dynamics at low
energies. The Schr{\"o}dinger equation is local in time, and the
interaction Hamiltonian describes an instantaneous interaction.
This is the main cause of infinities in the Hamiltonian formalism.
In Ref.[3] it has been shown that the use of the Feynman approach
to quantum theory in combination with the canonical approach
allows one to extend quantum dynamics to describe the evolution of
a system whose dynamics is generated by nonlocal in time
interaction.  A generalized quantum dynamics (GQD) developed in
this way has been shown to open new possibilities to resolve the
problem of the UV divergences in quantum field theory [3]. An
equation of motion has been derived as the most general dynamical
equation consistent with the current concepts of quantum theory.
Being equivalent to the Schr{\"o}dinger equation in the particular
case where interaction is instantaneous, this equation permits the
generalization to the case where the interaction operator is
nonlocal in time. Note that there is one-to-one correspondence
between nonlocality of interaction and the UV behavior of the
matrix elements of the evolution operator as a function of
momenta: The interaction operator can be nonlocal in time only in
the case where this behavior is "bad", i.e. in a local theory it
leads to the UV divergences. For this reason one can expect the
nucleon dynamics that follows from renormalization of an EFT to be
governed by the generalized dynamical equation with
nonlocal-in-time interaction operator. In the present paper this
problem is investigated by using the example of a toy model of the
separable NN potential which gives rise to logarithmic
singularities in the Born series. We show that the T matrix
obtained in Ref.[4] by dimensional regularization of this model
does not satisfy the Lippmann-Schwinger (LS) equation but
satisfies the generalized dynamical equation with a
nonlocal-in-time interaction operator.

In the GQD the following assumptions are used as basic postulates:

(i) The physical state of a system is represented by a vector
(properly by a ray) of a Hilbert space.

(ii) An observable A is represented by a Hermitian hypermaximal
operator $\alpha$. The eigenvalues $a_r$ of $\alpha$ give the
possible values of A. An eigenvector $|\varphi_r^{(s)}>$
corresponding to the eigenvalue $a_r$ represents a state in which
A has the value $a_r$. If the system is in the state $|\psi>,$ the
probability $P_r$ of finding the value $a_r$ for A, when a
measurement is performed, is given by
$$P_{r} = <\psi|P_{V_{r}} |\psi>= \sum_s |<\varphi_r^{(s)}|\psi>|^2, $$
where $P_{V_{r}}$ is the projection operator on the eigenmanifold
$V_r$ corresponding to $a_r,$ and the sum $\Sigma_s$ is taken over
a complete orthonormal set ${|\varphi_r^{(s)}>}$ (s=1,2,...) of
$V_r.$ The state of the system immediately after the observation
is described by the vector $P_{V_{r}}|\psi>.$

These  assumptions are the main assumptions on which quantum
theory is founded. In the canonical formalism they are used
together with the assumption that the time evolution of a state
vector is governed by the Schr{\"o}dinger equation. In the
formalism [3] this assumption is not used. Instead the assumptions
(i) and (ii) are used together with the following postulate.

(iii) The probability of an event is the absolute square of a
complex number called the probability amplitude. The joint
probability amplitude of a time-ordered sequence of events is
product of the separate probability amplitudes of each of these
events. The probability amplitude of an event which can happen in
several different ways is a sum of the probability amplitudes for
each of these ways.

The statements of the assumption (iii) express the well-known law
for the quantum-mechanical probabilities. Within the canonical
formalism this law is derived as one of the consequences of the
theory. However, in the Feynman approach this law is  used as a
basic postulate of the  theory.

It is also used the assumption that the time evolution of a
quantum system is described by the evolution equation
$|\Psi(t)>=U(t,t_0)|\Psi(t_0)>,$ where $U(t,t_0)$ is the unitary
evolution operator
\begin{equation}
U^{+}(t,t_0) U(t,t_0) = U(t,t_0) U^{+}(t,t_0) = {\bf 1},
\end{equation}
with the group property $ U(t,t') U(t',t_0) = U(t,t_0), \quad
U(t_0,t_0) ={\bf 1}. $ Here we use the interaction picture.

According to the assumption (iii), the probability amplitude of an
event which can happen in several different ways is a sum of
contributions from each alternative way. In particular, the
amplitude
 $<\psi_2| U(t,t_0)|\psi_1>$ can be represented as a sum
of contributions from all alternative ways of realization of the
corresponding evolution process. Dividing these alternatives in
different classes, we can then analyze such a probability
amplitude in different ways. For example, subprocesses with
definite instants of the beginning and  end of the interaction in
the system can be considered as such alternatives. In this way the
amplitude $<\psi_2|U(t,t_0)|\psi_1>$  can be written in the form
[3]
\begin{equation}
<\psi_2| U(t,t_0)|\psi_1> = <\psi_2|\psi_1> + \int_{t_0}^t dt_2
\int_{t_0}^{t_2} dt_1 <\psi_2|\tilde S(t_2,t_1)|\psi_1>,
\end{equation}
where $<\psi_2|\tilde S(t_2,t_1)|\psi_1>$ is the probability
amplitude that if at time $t_1$ the system was in the state
$|\psi_1>,$ then the interaction in the system will begin at time
$t_1$ and will end at  time $t_2,$ and at this time the system
will be in the state $|\psi_2>.$  In the case of an isolated
system the operator $\tilde S(t_2,t_1)$ can be represented in the
form $\tilde S(t_2,t_1) = exp(iH_0t_2) \tilde T(t_2-t_1) exp(-iH_0
t_1),$ $H_0$ being the free Hamiltonian [3]. As has been shown in
Ref.[3], for the evolution operator $U(t,t_0)$ given by (2) to be
unitary for any times $t_0$ and $t$, the operator $\tilde
S(t_2,t_1)$ must satisfy the following equation:
\begin{equation}
(t_2-t_1) \tilde S(t_2,t_1) = \int^{t_2}_{t_1} dt_4
\int^{t_4}_{t_1}dt_3 (t_4-t_3) \tilde S(t_2,t_4) \tilde
S(t_3,t_1).
\end{equation}
This equation allows one to obtain the operators $\tilde
S(t_2,t_1)$ for any $t_1$ and $t_2$, if the operators $\tilde
S(t'_2, t'_1)$ corresponding to infinitesimal duration times $\tau
= t'_2 -t'_1$ of interaction are known. It is natural to assume
that most of the contribution to the evolution operator in the
limit $t_2 \to t_1$ comes from the processes associated with the
fundamental interaction in the system under study. Denoting this
contribution by $H_{int}(t_2,t_1)$, we can write
\begin{equation}
\tilde{S}(t_2,t_1) \tend \limits_{t_2\rightarrow t_1}
H_{int}(t_2,t_1) + o(\tau^\varepsilon).
\end{equation}

Within the GQD the operator $H_{int}(t_2,t_1)$ plays the role
which the interaction Hamiltonian plays in the ordinary
formulation of quantum theory: It generates dynamics of a system.
Being a generalization of the interaction Hamiltonian, this
operator is called the generalized interaction operator. The
parameter $\varepsilon$ is determined by demanding that
$H_{int}(t_2,t_1)$ must be so close to the solution of Eq.(3) in
the limit $t_2\tend t_1$ that this equation has a unique solution
having the behavior (4) near the point $t_2=t_1$.

If  $H_{int}(t_2,t_1)$ is specified, Eq.(3) allows one to find the
operator $\tilde S(t_2,t_1).$ Formula (2) can then be used to
construct the evolution operator $U(t,t_0)$ and accordingly the
state vector $|\psi(t)> = |\psi(t_0)> +  \int_{t_0}^t dt_2
\int_{t_0}^{t_2} dt_1 \tilde S(t_2,t_1) |\psi(t_0)> $ at any time
$t.$ Thus Eq.(3) can be regarded as an equation of motion for
states of a quantum system. By using (2), the evolution operator
can be represented in the form [3]
$$<n_2|U(t,t_0)|n_1>=  <n_2|n_1>+ $$
\begin{equation}
+\frac{i}{2\pi} \int^\infty_{-\infty} dx
 \frac {exp[-i(z-E_{n_2})t] <n_2|T(z)|n_1> exp[i(z-E_{n_1})t_0]}
{(z-E_{n_2})(z-E_{n_1})},
\end{equation}
where $z=x+iy$,  and $y>0$, $n$ stands for the entire set of
discrete and continuous variables that characterize the system in
full, $|n>$ are the eigenvectors of the free Hamiltonian, and
$<n_2|T(z)|n_1>$ is defined by
\begin{equation}
<n_2|T(z)|n_1> = i \int_{0}^{\infty} d\tau exp(iz\tau) <n_2|\tilde
T(\tau)|n_1>.
\end{equation}
The generalized equation of motion (3) is equivalent to the
following equation for the T matrix [3]:
\begin{equation}
\frac{d <n_2|T(z)|n_1>}{dz} = -  \sum \limits_{n}
\frac{<n_2|T(z)|n><n|T(z)|n_1>}{(z-E_n)^2}.
\end{equation}
 Thus,
instead of solving Eq.(3), one can solve Eq.(7) for the T matrix.
The operator $\tilde T(\tau)$ and correspondingly the operator
$\tilde S(t_2,t_1)$ can then be obtained by using the Fourier
transform
\begin{equation}
<n_2|\tilde T(\tau)|n_1> = - \frac{i}{2\pi} \int
_{-\infty}^{\infty}dx exp(-iz\tau) <n_2|T(z)|n_1>.
\end{equation}
 At the same time, the T matrix can be directly used for constructing the
evolution operator. According to (4) and (8), the operator $T(z)$
has the following asymptotic behavior for $|z| \tend \infty:$
\begin{equation}
<n_2|T(z)|n_1> \tend \limits_{|z| \tend \infty} i \int_0^{\infty}
d\tau exp(iz \tau)<n_2| H^{(s)}_{int}(\tau) |n_1> +
o(|z|^{-\beta}) ,
\end{equation}
where $\beta=1+\epsilon,$ and $H^{(s)}_{int}(t_2-t_1) =
exp(-iH_0t_2) H_{int}(t_2,t_1) exp(iH_0t_1)$ is the interaction
operator in the Schr{\"o}dinger picture. As has been shown in
Ref.[3], the dynamics governed by Eq.(3) is equivalent to the
Hamiltonian dynamics in the case where the generalized interaction
operator is of the form
\begin{equation}
 H_{int}(t_2,t_1) = - 2i \delta(t_2-t_1)
 H_{I}(t_1) ,
\end{equation}
$H_{I}(t_1)$ being the interaction Hamiltonian in the interaction
picture. In this case the state vector $|\psi(t)>$ given by (2)
satisfies the Schr{\"o}dinger equation $
  \frac {d |\psi(t)>}{d t} = -iH_I(t)|\psi(t)>.
$ The delta function $\delta(\tau)$ in (10) emphasizes the fact
that in this case the fundamental interaction is instantaneous. At
the same time, Eq.(3) permits the generalization to the case where
the interaction generating the dynamics of a system is nonlocal in
time [3,5].

As has been shown [3,5], there is one-to-one correspondence
between nonlocality of interaction and the UV behavior of the
matrix elements of the evolution operator as a function of momenta
of particles: The interaction operator can be nonlocal in time
only in the case where this behavior is "bad", i.e. in a local
theory it results in UV divergences. In Ref.[6] it has been shown
that after renormalization the dynamics of the three-dimensional
theory of a neutral scalar field interacting through a $\varphi^4$
coupling is governed by the generalized dynamical equation (3)
with a nonlocal-in-time interaction operator. This gives reason to
expect that after renormalization the dynamics of an EFT is also
governed by this equation with a nonlocal interaction operator.
Let us now consider this problem on a toy model of the separable
NN interaction. We will use the c.m.s., and denote the relative
momentum of two nucleons by ${\bf p}$ and the reduced mass by
$\mu.$ In this model the generalized interaction operator should
be of the form
\begin{equation}
<{\bf p}_2| H^{(s)}_{int}(\tau) |{\bf p}_1> = \psi^*({\bf p}_2)
\psi({\bf p}_1) f(\tau),
\end{equation}
where $f(\tau)$ is some function of duration time of interaction
$\tau$. Let the form factor $\psi ({\bf p})$ has the following UV
behavior:
\begin{equation}
\psi({\bf p}) \sim |{\bf p}|^{-\alpha}, \quad {(|{\bf p}| \tend
\infty).}
\end{equation}
Assume, for example,that $\psi({\bf p}) = |{\bf
p}|^{-\alpha}+g({\bf {p}}),$ and in the limit $|{\bf p}| \tend
\infty$ the function $g({\bf {p}})$ satisfies the estimate $g({\bf
{p}})=o(|{\bf {p}}|^{-\delta})$, where $\delta>\alpha,$
$\delta>\frac{3}{2}.$ In this case, the problem can be easily
solved by using Eq.(7). Representing $<{\bf p}_2| T(z) |{\bf
p}_1>$ in the form $ <{\bf p}_2| T(z)|{\bf p}_1> = \psi^* ({\bf
p}_2)\psi ({\bf p}_1) t(z), $ from (7), we get the equation
\begin{equation}
\frac {dt(z)}{dz} = -t^2(z) \int d^3k \frac {|\psi ({\bf k})|^2}
{(z-E_k)^2}
\end{equation}
with the asymptotic condition
\begin{equation}
t(z)  \tend \limits_{|z| \tend \infty} f_1(z) + o(|z|^{-\beta}),
\end{equation}
$f_1(z)= i \int_0^{\infty} d\tau exp(iz\tau) f(\tau), $ and $E_k =
\frac {k^2}{2 \mu}.$ The solution of Eq.(13) with the initial
condition $t(a)=g_a,$ where $a \in (-\infty,0),$ is
\begin{equation}
t(z) = g_a \left(1 +(z-a) g_a
 \int d^3k \frac {|\psi ({\bf k})|^2}
{(z-E_k)(a-E_k)} \right)^{-1}.
\end{equation}
In the case $\alpha >\frac{1}{2}$, the function $t(z)$ tends to a
constant as $z \tend -\infty$: $ t(z)  \tend \limits_{z \tend
-\infty} \lambda. $ Thus in this case the function $f_1(z)$ must
tend to $\lambda$ as $z \tend -\infty.$ From this it follows that
only possible form of the function $f(\tau)$ is $ f(\tau) = -2i
\lambda \delta(\tau) + f^{\prime}(\tau), $ where the function
$f^{\prime}(\tau)$ has no such a singularity at the point $\tau=0$
as the delta function. In this case  the generalized interaction
operator is of the form (10), and hence the dynamics generated by
this operator is equivalent to the dynamics governed by the
Schr{\"o}dinger equation with the separable potential $\lambda
\psi^*({\bf p}_2) \psi({\bf p}_1).$ In this case the LS equation
follows from Eq.(7), and our model is equivalent to the ordinary
separable-potential model.

Ordinary quantum mechanics does not permit the extension of the
above model to the case $\alpha \leq \frac{1}{2}$, because of the
UV divergences. The GQD allows one to extend this model to the
case $-\frac{1}{2} < \alpha \leq\frac{1}{2}.$ In Ref.[5] such an
extended model has been constructed for $-\frac{1}{2} < \alpha
<\frac{1}{2}$. Let us now consider the case $\alpha=\frac{1}{2}$.
In this case the UV behavior of the form factor $\psi({\bf p})$
gives rise to the logarithmic singularities in the Born series. At
first we have to determine the class of the functions  $f_1(z)$
and the value of $\beta$ for which Eq.(13) has a unique solution
having the asymptotic behavior (14). In the case $\alpha
=\frac{1}{2},$ the function $t(z)$ given by (15) has the following
behavior for $|z| \tend \infty:$
\begin{equation}
t(z)  \tend \limits_{|z| \tend \infty}  b_1 \ln^{-1}(-z)+ b_2
\ln^{-2}(-z) + o(\ln^{-2}(-z)),
\end{equation}
where $b_1 =- (4\pi \mu)^{-1}$ and $b_2= b_1 \ln(-a)
-b_1^2(M(a)+g_a^{-1})$ with
\begin{equation}
M(a) = \int \frac {|\psi({\bf k})|^2-|{\bf {k}}|^{-2\alpha}}
{a-E_k}d^3k .
\end{equation}
The parameter $b_1$ does not depend on $g_a.$ This means that all
solutions of Eq.(13) have the same leading term  in (16), and only
the second term distinguishes the different solutions of this
equation. Thus, in order to obtain a unique solution of Eq.(13),
we must specify the first two terms in the asymptotic behavior of
$t(z)$ for $|z| \tend  \infty.$ From this it follows that the
functions $f_1(z)$ must be of the form
\begin{equation}
f_1(z) = b_1\ln^{-1}(-z)+ b_2 \ln^{-2}(-z) ,
\end{equation}
and $\beta=2 \alpha -1.$ By using (8) and the definition of
$f_1(z)$, we can then obtain the functions $f(\tau)$ and hence the
generalized interaction operator
\begin{equation}
 <{\bf p}_2| {H}^{(s)}_{int}(\tau)|{\bf p}_1> = -\frac{i}{2\pi}
 \psi^* ({\bf p}_2) \psi ({\bf
 p}_1)\int_{-\infty}^{\infty}dx
 \exp(-iz\tau)\left(\frac{b_1}{\ln(-z)}+\frac{b_2}{ \ln^{2}(-z)}\right),
\end{equation}
where  $z=x+iy,$ and $y>0.$ Thus, in the case of the UV behavior
of the form factors corresponding to $\alpha=\frac{1}{2}$, the
interaction operator $H^{(s)}_{int}(\tau)$ necessarily has the
form (19), where for given $\psi({\bf p})$ only the parameter
$b_2$ is free. The solution of Eq.(7) with this interaction
operator is of the form
\begin{equation}
<{\bf p}_2| T(z)|{\bf p}_1> = N(z) \psi^* ({\bf p}_2)\psi ({\bf
p}_1),
\end{equation}
where $ N(z)=b_1^2\left(-b_2+b_1\ln(-z)+M(z)b_1^2\right)^{-1}. $

In Refs.[5] the model for $-\frac{1}{2}<\alpha<\frac{1}{2}$ (in
this case the interaction operator is nonlocal in time) was used
for describing the NN interaction at low energies. The motivation
to use such a model for the parameterization of the NN forces is
the fact that due to the quark and gluon degrees of freedom the NN
interaction must be nonlocal in time. However, because of the
separation of scales the system of hadrons should be regarded as a
closed system. From this it follows that the dynamics of such a
system is governed by Eq.(3) with nonlocal-in-time interaction
operator. Note that the UV divergences of EFT's are the price for
ignoring the quark and gluon degrees of freedom in describing
hadron dynamics: EFT's are local theories, despite the existence
of the external quark and gluon  degrees of freedom. However,
renormalization of EFT's gives rise to the fact that these
theories become nonlocal. Below this will be shown by using the
example of our toy model.

As we have stated, the interaction operator (19) contains only one
free parameter $b_2$. However, if there is a bound state in the
channel under study, then the parameter $b_2$ is completely
determined by demanding that the T matrix has  the pole at the
bound-state energy. For example, in the ${}^3S_1$ channel the T
matrix has a pole at energy $E_B=-2.2246$MeV. This means  that
$\left[t(E_B)\right]^{-1}=0$,  and, by putting $a=E_B$ in Eq.(15),
we get
\begin{equation}
\left [ t(z)\right ]^{-1}=(z-E_B)\int d^3 k\frac{|\psi({\bf
{k}})|^2}{(z-E_k) (E_B-E_k)}.
\end{equation}
In this case $b_2=b_1\ln(-E_B)-b_1^2M(E_B)$. Let us now show that
renormalization of the LS equation with the separable potential
leading to a logarithmic singularities produces the same T matrix.
In [4] the problem of renormalization of the LS equation was
considered by using the example of the separable potential $<{\bf
p}_2|V|{\bf p}_1>=\lambda\psi^*({\bf p}_2)\psi({\bf p}_1)$ with
$\psi({\bf {p}})=(d^2+{\bf {p}}^2)^{-\frac{1}{4}}$. The
corresponding T matrix is of the form $<{\bf p}_2|T(z)|{\bf
p}_1>=\psi^*({\bf p}_2)\psi({\bf p}_1)t(z)$ with
\begin{equation}
t(z)=\left (\lambda^{-1} -J(z)\right)^{-1},
\end{equation}
where the integral $ J(z)=\int d^3k\frac{|\psi({\bf
{k}})|^2}{z-E_k}$ has a logarithmic divergence. By using the
dimensional regularization, one can get
\begin{equation}
\left [t_{\varepsilon}(z)\right ]^{-1}=
\lambda_{\varepsilon}^{-1}-J_{\varepsilon}(z),
\end{equation}
where $J_{\varepsilon}(z)=\int d^{3-\varepsilon}k\frac {|\psi({\bf
{k}})|^2}{z-E_k}.$ The strength of the potential
$\lambda_{\varepsilon}$ is adjusted to give the correct
bound-state energy:
$\lambda_{\varepsilon}^{-1}=J_{\varepsilon}(E_B).$ In this way we
get
\begin{equation}
[t_{\varepsilon}(z)]^{-1}=J_{\varepsilon}(E_B)-J_{\varepsilon}(z)=
(z-E_B)\int d^{3-\varepsilon}k\frac {|\psi({\bf
{k}})|^2}{(z-E_k)(E_B-E_k)}.
\end{equation}
The right-hand side of (29) is well-behaved in the limit
$\varepsilon\rightarrow 0$. It is easy to see that, taking this
limit, we get the expression (22) for $t(z)$. Thus renormalization
of the LS equation with the above singular potential leads to the
T matrix we have obtained by solving Eq.(7) with nonlocal-in-time
interaction operator. This T matrix satisfies the generalized
dynamical equation (7), but does not satisfy the LS equation.
Correspondingly the Schr{\"o}dinger equation is not  valid in this
case. The strength of the potential $\lambda_\varepsilon$ tends to
zero as $\varepsilon\rightarrow 0$, and consequently the
renormalized interaction Hamiltonian is also tend to zero. Thus
despite the fact that the dynamics of each theory corresponding to
the dimension ${\cal D}=3-\varepsilon$ is Hamiltonian dynamics for
every $\varepsilon>0$, in the limiting case ${\cal D}=3$ we go
beyond Hamiltonian dynamics. The dynamics of the renormalized
theory is governed by the generalized equation of motion (3) with
nonlocal-in-time interaction operator which in the case of the
model under study is given by (19).

To conclude, by using the model of the separable NN potential
which gives rise to logarithmic singularities in the Born series,
we have demonstrated that after renormalization the dynamics of a
nucleon system is governed by the generalized dynamical equation
with a nonlocal-in-time interaction operator.  This gives reason
to hope that the use of the GQD and a parameterization of the NN
forces like (19) can open new possibilities for applying the EFT
approach to the description of low-energy nucleon dynamics. By
using an EFT, one can construct the generalized interaction
operator consistent with the symmetries of QCD. This operator can
then be used in Eq.(3) for describing nucleon dynamics. Note that
in this case regularization and renormalization are needed only on
the stage of determining the two-nucleon interaction operator.
After this one does not face the problem of UV divergences.
\newpage
\section*{References}

\begin{enumerate}

\item[{[1]}]
S. Weinberg, Physica A {\bf 96} (1979) 327; E. Witten, Nucl. Phys.
B {\bf 122} (1977) 109; H. Georgi, Annu.Rev.Nucl.Part.Sci. {\bf
43} (1993) 209; J. Polchinski, Proceedings of Recent Directions in
Particle Theory, TASI-92 (1992) 235; A.V. Manohar, Effective Field
Theories, hep-ph/9508245; D.B. Kaplan, Effective Field Theories,
nucl-th/9506035.
\item[{[2]}]
T. Mehen and I.W. Stewart, Phys.Rev. C {\bf 59} (1998) 2365; C.
Ordonez, and U. van Kolck, Phys. Lett. B {\bf 291} (1992) 459; U.
van Kolck, Phys. Rev. C {\bf 49} (1994) 2932; D.B. Kaplan, M.J.
Savage, and M.B. Wise, Phys.Lett. B {\bf 424} (1998) 390;
Nucl.Phys. B {\bf 534} (1998) 329.
\item[{[3]}]
R.Kh. Gainutdinov, J. Phys. A: Math. Gen. {\bf 32} (1999) 5657.
\item[{[4]}]
D.R. Phillips, I.R. Afnan, and A.G. Henry-Edwards, Phys.Rev. C.
{\bf 61} (2000) 044002.
\item[{[5]}]
R. Kh. Gainutdinov and  A.A.Mutygullina,
 Yad. Fiz. {\bf 60} (1997) 938 [Physics of Atomic Nuclei,
{\bf 60} (1997) 841]; {\bf 62} (1999) 2061 [Physics of Atomic
Nuclei, {\bf 62} (1999) 1905].
\item[{[6]}]
R. Kh. Gainutdinov, hep-th/0107139.
\end{enumerate}
\end{document}